\documentstyle[preprint,aps]{revtex}
\begin{document}
\draft
\vskip 2cm
\preprint{SNUTP 93-71}
\title{\bf Factorization in graviton interactions}
\author{S. Y. Choi, J. S. Shim, and H. S. Song}
\vskip 1cm
\address{Center for Theoretical Physics and Department of Physics,\\
         Seoul National University, Seoul 151-742, Korea}
\maketitle
\vskip 5cm
\begin{abstract}
The study of factorization in the linearized gravity is extended
to the graviton scattering processes with a massive scalar
particle, with a massless vector boson and also with a graviton.
Every transition amplitude is shown to be completely factorized and
the physical implications of their common factors are discussed.
\end{abstract}
\vskip 0.5cm

%
%
Gravitational gauge invariance and graviton transversality
force transition amplitudes of four-body graviton interactions
to be factorized. Yang-Mills theories\cite{Zhu} also exhibit a
similar factorization property.

In our previous work, we have shown that the transition amplitudes
of $ge\rightarrow\gamma e$\cite{Choi1} and
$ge\rightarrow ge$\cite{Choi2} processes are completely factorized
into a kinematical factor, a fermion-QED Compton scattering form,
and another gauge invariant term.
In particular, the transition amplitude of graviton scattering
with a massive fermion reads
\begin{eqnarray}
{\cal M}_{gf}&=&-\frac{\kappa^2}{2}
            F\left[(\epsilon_1\!\cdot\!\epsilon_2)
            +\frac{(p_1\!\cdot\!\epsilon_1)
            (p_2\!\cdot\!\epsilon_2)}{(p_1\!\cdot\! k_1)}
            +\frac{(p_2\!\cdot\!\epsilon_1)
            (p_1\!\cdot\!\epsilon_2)}{(p_1\!\cdot\! k_2)}
     \right]\nonumber\\
     &\times&\hskip 0.2cm\bar{u}(p_2)
     \left[\frac{\not\!{\epsilon}_2(\not\!{p}_1+\not\!{k}_1+m)
     \!\!\not\!{\epsilon}_1}{2(p_1\!\cdot\! k_1)}
     +\frac{\not\!{\epsilon}_1(\not\!{p}_1+\not\!{k}_2+m)\!\!\not\!
     {\epsilon}_2}{2(p_1\!\cdot\! k_2)}\right]u(p_1),
 \label{fac1}
\end{eqnarray}
where all particles are assumed to be incoming,
the $\epsilon_1^\mu\epsilon_1^\nu$($\epsilon_2^\mu\epsilon_2^\nu$)
and $k_1^\mu$($k_2^\mu$) are the wave tensor and four-momentum of
the initial(final) graviton, and $p_1$($p_2$) is the four-momentum
of the initial(final) fermion, respectively. The coupling is
denoted by $\kappa=\sqrt{32\pi G_N}$ and the overall kinematical
factor $F$ is given by
\begin{eqnarray}
F=\frac{(p_1\!\cdot\! k_1)(p_1\!\cdot\! k_2)}{(k_1\!\cdot\! k_2)}.
\end{eqnarray}
The first bracket term in Eq.\ (\ref{fac1}) is none other than the
photon-scalar scattering amplitude and the second one is the
photon-fermion  scattering amplitude.

In this paper, we extend the study of factorization in the
linearized gravity to the graviton scattering processes with a
massive scalar, a massless vector boson, and a graviton itself.

Let us now introduce a manifestly gauge invariant four-vector
$\tilde{\epsilon_i}$($i=1,2$) defined as
\begin{eqnarray}
\tilde{\epsilon_i}=\epsilon_i
       -\frac{(p_1\!\cdot\!\epsilon_i)}{(p_1\!\cdot\! k_i)}k_i.
 \label{fac3}
\end{eqnarray}
As a result, the expression of photon-scalar scattering amplitude
is converted into a greatly simplified form:
\begin{eqnarray}
 \epsilon_1\!\cdot\!\epsilon_2
+\frac{(p_1\!\cdot\!\epsilon_1)
 (p_2\!\cdot\!\epsilon_2)}{(p_1\!\cdot\! k_1)}
+\frac{(p_2\!\cdot\!\epsilon_1)
 (p_1\!\cdot\!\epsilon_2)}{(p_1\!\cdot\! k_2)}
     =\tilde{\epsilon_1}\!\cdot\!\tilde{\epsilon_2}.
 \label{fac4}
\end{eqnarray}
Along with this simplification, gravitational gauge invariance
and graviton transversality render the transition amplitude
${\cal M}_{gs}$ of graviton-scalar scattering and the transition
amplitude ${\cal M}_{gv}$ of graviton scattering with a massless
vector boson completely factorized and simple:
\begin{eqnarray}
{\cal M}_{gs}&=&\frac{\kappa^2}{2}F
   \left[\tilde{\epsilon_1}\!\cdot\!\tilde{\epsilon_2}\right]^2,
 \label{fac2}
\end{eqnarray}
and
\begin{eqnarray}
{\cal M}_{gv}&=&-\frac{\kappa^2}{2}F
      [\tilde{\epsilon_1}\!\cdot\!\tilde{\epsilon_2}]
      \left[(\tilde{\epsilon_1}\!\cdot\!\tilde{\epsilon_2})
      (\tilde{\varepsilon_1}\!\cdot\!\tilde{\varepsilon_2})
      \!+(k_1\!\cdot\! k_2)\!
      \left\{\frac{(\tilde{\epsilon_1}
      \!\cdot\!\tilde{\varepsilon_1})
      (\tilde{\epsilon_2}\!\cdot\!
      \tilde{\varepsilon_2})}{p_1\!\cdot\! k_1}
      +\frac{(\tilde{\epsilon_1}\!\cdot\!\tilde{\varepsilon_2})
      (\tilde{\epsilon_2}\!\cdot\!
      \tilde{\varepsilon_1})}{p_1\!\cdot\! k_2}
      \right\}\right.\nonumber\\
     &&\left. \hskip 2.9cm
      +\frac{1}{p_1\!\cdot\! k_2}
      \{(\tilde{\epsilon_2}\!\cdot\!\tilde{\varepsilon_1})
      (p_2\!\cdot\!\tilde{\epsilon_1})
      (k_2\!\cdot\!\tilde{\varepsilon_2})
      +(\tilde{\epsilon_1}\!\cdot\!\tilde{\varepsilon_2})
      (p_2\!\cdot\!\tilde{\epsilon_2})
      (k_2\!\cdot\!\tilde{\varepsilon_1})
     \right.\nonumber\\
     &&\left. \hskip 2.9cm
      -(\tilde{\epsilon_2}\!\cdot\!\tilde{\varepsilon_2})
      (p_2\!\cdot\!\tilde{\epsilon_1})
      (k_2\!\cdot\!\tilde{\varepsilon_1})\}
      +\frac{1}{p_1\!\cdot\! k_1}
      (\tilde{\epsilon_1}\!\cdot\!\tilde{\varepsilon_1})
      (p_2\!\cdot\!\tilde{\epsilon_2})
      (k_2\!\cdot\!\tilde{\varepsilon_2})\right],
 \label{fac5}
\end{eqnarray}
where $p_1$($p_2$) and $\varepsilon_1$($\varepsilon_2$) are
the four-momentum and wave vector of the initial(final) massless
vector boson, respectively, and $\tilde{\varepsilon_i} (i=1,2)$
is defined in a gauge invariant form as
\begin{eqnarray}
\tilde{\varepsilon_i}=\varepsilon_i
     -\frac{(k_1\!\cdot\! \varepsilon_i)}{(k_1\!\cdot\! p_i)} p_i.
  \label{fac6}
\end{eqnarray}
The second bracket term of Eq.\ (\ref{fac5}) corresponds to
the transition amplitude of photon scattering with a massless
charged vector boson. Let us employ a symbol ${\cal M}_{\gamma v}$
for the amplitude. Likewise, the transition amplitude
of graviton-graviton scattering \cite{Sannan} turns out to be factorized
as
\begin{eqnarray}
{\cal M}_{gg}=\frac{\kappa^2}{2}
              F\left[{\cal M}_{\gamma v}\right]^2.
  \label{fac8}
\end{eqnarray}
Here $p_1(p_2)$ and $\varepsilon_1^\mu \varepsilon_1^\nu
(\varepsilon_2^\mu \varepsilon_2^\nu)$ are the four-momentum and
wave tensor of another initial(final) graviton, respectively.
While the form of Eq.\ (\ref{fac4}) is independent of the choice of
$\tilde{\epsilon_i}$, the photon-vector boson scattering amplitude
${\cal M}_{\gamma v}$  can be changed if the $\tilde{\epsilon_i}$
and $\tilde{\varepsilon_i}$ are defined in a different way.
Nevertheless, the transition amplitude ${\cal M}_{gg}$ satisfies
Bose and crossing symmetries. The completely symmetric expression of
${\cal M}_{gg}$ can be found in Ref. \cite{Sannan}.

Several comments are in order.
\begin{itemize}
\item[{(a)}] The $\tilde{\epsilon_i}$ and $\tilde{\varepsilon_i}$
             are manifestly gauge invariant and so are all
             four-body transition amplitudes.
\item[{(b)}] The helicity amplitudes\cite{Cho,Berends} for the
             graviton-scalar,
             graviton-photon, and graviton-graviton scattering
             processes can be immediately obtained from our
             corresponding factorized transition amplitudes.
\item[{(c)}] The expression $[\tilde{\epsilon_1}\!\cdot\!
             \tilde{\epsilon_2}]$ always appears as a common factor
             in the graviton scattering with matter fields such as
             a massive scalar, a massive fermion, and a massless
             vector boson, while the ${\cal M}_{\gamma v}$ is a
             common factor in the graviton scattering processes
             with a massless vector boson and a graviton itself.
\end{itemize}

To conclude, every four-body graviton scattering amplitude is
completely factorized and the introduction of manifestly gauge
invariant four-vectors $\tilde{\epsilon_i}$ and
$\tilde{\varepsilon_i}$($i=1,2$) renders the amplitude very simple.
These two simplifications justify why, with all the very complicated
three-graviton and four-graviton vertices\cite{Berends,DeWitt},
the final form of transition amplitudes is so simple.
The factorization property looks so generic that its more intensive
and extensive investigation is expected to provide us with some
crucial clues for the unification of gravity with other interactions.
\vskip 3cm
\section*{Acknowledgments}

The work was supported in part by the Korea Science and Engineering
Foundation through the SRC program and in part by the Korean
Ministry of Education

\end{document}